# How are academic age, productivity and collaboration related to citing behavior of researchers?

*Staša Milojević*

Accepted for publication in PLoS One

*School of Library and Information Science, Indiana University, Bloomington, IN, USA*

smilojev@indiana.edu

## ABSTRACT

References are an essential component of research articles and therefore of scientific communication. In this study we investigate referencing (citing) behavior in five diverse fields (astronomy, mathematics, robotics, ecology and economics) based on 213,756 core journal articles. At the macro level we find: (a) a steady increase in the number of references per article over the period studied (50 years), which in some fields is due to a higher rate of usage, while in others reflects longer articles and (b) an increase in all fields in the fraction of older, foundational references since the 1980s, with no obvious change in citing patterns associated with the introduction of the Internet. At the meso level we explore current (2006-2010) referencing behavior of different categories of authors (21,562 total) within each field, based on their academic age, productivity and collaborative practices. Contrary to some previous findings and expectations we find that senior researchers use references at the same rate as their junior colleagues, with similar rates of re-citation (use of same references in multiple papers). High Modified Price Index (MPI, which measures the speed of the research front more accurately than the traditional Price Index) of senior authors indicates that their research has the similar cutting-edge aspect as that of their younger colleagues. In all fields both the productive researchers and especially those who collaborate more use a significantly lower fraction of foundational references and have much higher MPI and lower re-citation rates, i.e., they are the ones pushing the research front regardless of researcher age. This paper introduces improved bibliometric methods to measure the speed of the research front, disambiguate lead authors in co-authored papers and decouple measures of productivity and collaboration.

## Introduction

Communication of scientific results is an integral part of modern science, making scientific results "visible" to other scientists and to society as a whole. Through the years science has perfected acceptable genres and discourses of disseminating its results [1,2]. Journals and journal articles serve as primary vehicles not only for disseminating the findings, but also for reinforcing the common paradigms of scientific fields and disciplines. Each journal article is written for a particular audience and is adhering to certain rhetorical devices in order to establish trust and authority. Modern science is based on trust [3].

One important way of establishing trust and authority is through lineage of methods, theories and problems used. That lineage is manifested in the lists of bibliographic references[1], which have become a staple of every scientific article since the end of the 19th century [4,5]. Thus, referencing, in addition to providing a context for a study, is a necessary device to persuade editors, referees, and ultimately the scientific audience of the study's credibility. The references themselves have undergone a transformation from mentions of authors and their work in the text to footnotes. Further transformation has occurred from footnotes to endnotes [1].

References have been exploited extensively to evaluate and describe science and trends in its development (especially since the development of Citation Indexes in the 1960s and 1970s). Research efforts that focus on references are primarily of two types: (a) studies that try to develop theories of citation by analyzing why and how people use references [e.g., 6,7-11] and (b) studies that explore the development of science and which use references to study moving research fronts, aging and obsolescence of scientific literature [12-18]. Less attention has been paid to authors' referencing *behavior* itself, especially using large bibliographic data sets, which forms the focus of this study. Thus we are interested in the act of citing and not its consequences, performing what Ajiferuke and colleagues call "citer-centered analysis" [19].

We distinguish three levels at which one can study referencing behavior: macro, meso and micro. At the macro, or global level, the units of analysis are entire fields, or even multiple fields, either at some fixed time or through time. Macro studies have recognized that different disciplines will *on average* have very different referencing practices that result in different characteristics of references [20,21]. These differences are the result of the traditions and practices in a given field and to a large degree reflect the fact that "science is a social process" [22]. Furthermore, the referencing practices evolve, again on a global level, usually on the time scales of decades, but sometimes faster [14]. In the first part of the Results section we explore the macro characteristics of the references in five fields that are chosen for this study, setting the stage for the analysis at the meso level. At the opposite end of the spectrum many studies approach referencing behavior at the micro level, i.e., as a deeply personal act done by individual authors. Such studies have often focused on the nature and the context of usage of references. For example, whether references are perfunctory or are essential to the paper [23], whether researchers cite to give credit [7] or to persuade [6], whether references are given in positive or negative connotations [10], or even whether references can be a product of deliberate "gaming" of a system [24]. Such micro-level behavior, although interesting, does not address how authors cite as a group. Namely, science is not only a personal, but also a communal activity. Our study fills the space between macro and micro studies because it focuses on the citing behavior of different *categories* of authors currently publishing. In order to better understand which meso characteristics are important for the scientific process in general, we explore the referencing characteristics of authors in five diverse scientific fields noting trends that transcend individual fields.

We explore the referencing behavior of authors classified according to the following characteristics: academic "age," productivity, and collaborative activity. Biological age of a researcher has been considered an important factor when it comes to creativity, productivity, and collaborative practices [25].

---

[1] References and citations are two sides of the same phenomenon. Price was among the first to differentiate between the two. In this paper we use the convention he proposed: "if Paper R contains a bibliographic footnote using and describing Paper C, then R contains a *reference* to C, and C has a *citation* from R" [4]. The current paper focuses entirely on references, and not on citations. Note, however, that when used as verbs (to *reference* and to *cite*) the two are practically synonymous.

Age has also been considered a factor when it comes to referencing behavior [12,13]. However, we are more interested in the cohorts of scientists who are at the same stage in their academic careers, having the same "academic" age, regardless of their biological age. Thus, we explore whether researcher seniority is correlated with certain properties of his/her references and referencing behavior. The reasons senior scientists might have different practices can be manifold. These researchers were enculturated into the practices of the field at a different time. Therefore, even if they are keeping pace with new developments in science, their referencing practices may contain "remnants" of those older times. Some studies have suggested that older authors tend to read less, and thus refer to older literature, which they recall from earlier times [26]. Because they have more experience, older authors may also be more selective when it comes to what they cite, leading to a smaller number of references. In this study we explore if such trends with academic age are actually present among the authors who publish in core journals in their fields.

The second characteristic of authors we are interested in is their current productivity. Scientific productivity, especially evidenced by a large number of publications, has become a holy grail of sorts for many scientists. Productivity has been closely tied to rewards in science. It has also been related to high quality of work [27]. Do more productive authors tend to have different referencing practices and strategies? Do prolific authors tend to cut down on the number of references (as suggested by [26]), or, does their increased productivity expose them to a larger body of literature that they cite? Our study will provide an answer to this question.

One of the most notable changes in science is a trend toward team science. Both the number of coauthored papers and the number of coauthors on a single paper have been increasing [28,29]. Collaboration in research is considered essential for tackling problems of present-day science and society. The above studies have also found that teams produce higher impact research. The increased importance of collaboration raises the question of whether scientists who tend to collaborate more have different referencing behaviors. Note that certain definitions of productivity make it automatically correlated to the level of collaboration, so in this study we will define measures which are independent from each other.

To be able to compare the characteristics of scientists with the appropriate reference group, we will follow each characteristic for each discipline separately.

Guided by the previous studies, we identify the following characteristics of references and referencing behavior as relevant: the number and age of references and the instances of re-citation. As pointed out, these characteristics have been studied primarily at a macro level, with special emphasis on trends over time. Price [30] was the first to write about the exponential growth in the number of references per paper. This observation has led to other studies of growth [e.g., 31]. All the studies agree that reference lists are getting longer. Age of references has been considered an important indicator for the vitality of a discipline and the pace of its advancement – the moving research front [32]. There has been a sense that the "younger" (more contemporaneous) the references are, the faster the advancement of science. Researchers have come up with a number of ways to measure the age: the simple average age, the "immediacy factor" [33], the "citation half-life" [34], and the "Price Index" [32], to name a few. We will use some of these and also introduce new or modified measures. One naively expects that the exponential growth of science will automatically lead to "younger" references. What is observed is the reverse [14] – the references are getting older. Part of the trend is a simple mathematical consequence of the exponential growth of literature with finite beginning [35]. Finally, we will study authors' *re-citation* practices, first suggested by White [36]. Ajiferuke, Lu and Wolfram [19] identified two types of re-citation: (a) at the author level (the same author cited in different articles of the same citing author) and (b) at the publication level (the same publication cited in different articles by the same citing author). For White, re-citation reveals authors' intellectual histories. Previous studies of re-citation practices focused on detailed analysis of *oeuvres* of individual researchers [36,37]. Here we present for the first time the analysis of re-citation at the meso level, for categories of different authors.

To summarize, the primary objective of the current study is to establish, for authors belonging to five disciplinary affiliations, the relationships between the characteristics of these authors (academic age, productivity, and collaboration) and their referencing behavior at the present time, as manifested through the number and age of references and re-citation practices. The secondary goal, setting the stage for the primary objective, is to describe the trends of these characteristics of references over the last half century, thereby establishing the macro-level properties of the references. Thus, this paper goes beyond the motivations and behavior of a single author; to use Small's [38] words, it moves from "the author-centric to the community-centric perspective".

## Methods

We analyze referencing practices in five diverse scientific fields ranging from classical to relatively new disciplines:

- Astronomy (AST; physical science; classical)
- Mathematics (MAT; mathematical science; classical)
- Robotics (ROB; technical/applied science; interdisciplinary and new)
- Ecology (ECL; life science; semi-classical)
- Economics (ECN; social science; classical)

The hope is that by examining diverse fields we will be able to sample various referencing practices and get a sense of which factors are discipline-specific and which are more generally related to the scientific process.

### Data

For each field we selected up to 10 *core* journals. These are journals that are well established and usually have high impacts in their field. In addition, core journals usually publish a large fraction of the original research in a particular discipline. Finally, and what is most important for our study, they allow for "coordination of communication and access to reputation, … knowledge interchange and creation" [39] making them good representatives of that field and its practices in general. These are also journals in which most active researchers will be publishing at some time in their careers. For our study it is not at all necessary to include the majority of journals in a given field, as the characteristics that we are interested in will be expressed in the core journals, as long as they contain a statistically large number of articles. Researchers who, for whatever reason, do not publish in core journals of their fields will not be encompassed by our study. We will discuss later on whether the absence of such authors affects the results.

We used a number of studies to identify the core journals for each discipline. For astronomy, [40] identified a list of core journals; for mathematics, our starting list came from journals identified by [41]; Sabanovic (personal communication) suggested the list for robotics; the list used for ecology came from [42] and, finally, the economics journals were identified by [43]. For these 33 journals (Table 1) we downloaded bibliographic data from Thomson Reuters' Web of Science database covering the 50-year interval between 1961 and 2010. We kept only data on research articles by selecting publications classified as "article" or "conference paper". In several cases in which the journal changed its title, we collected data corresponding to all predecessor titles. In total, our data set contained records for 213,756 articles. We checked the mean number of references per article in each journal to determine whether there were any journals with "special" referencing practices. We found that *Bulletin of the American Mathematical Society* (BAMS) contains almost twice as many references as other mathematics journals. After examining the journal we realized that BAMS publishes primarily explanatory papers, not original research articles. Because genre is one of the main determinants of referencing behavior, we removed this journal from the analysis. The total number of articles published in the 2006-2010 period is 45,043. Table 1 shows the breakdown by journal.

## Characteristics of authors

Our study did not follow the behavior of individual scientists through their careers, but rather focused on their collective behavior during the *current* period, which we defined as the five years from 2006 to 2010. Five years is sufficiently long enough to establish patterns of productivity and collaboration for most researchers, yet not too long to be affected by gradual changes in practice.

Referencing behavior was studied for all authors who at some point in what we call the current period were sole authors or *lead* authors on multiple-author papers. The inclusion of multiple-author papers was motivated by the model of authorship in which the lead author is primarily responsible for the published work, and therefore carries most weight in the selection of references. The identification of the lead author is not straightforward. The simplest assumption is that the first author is also the lead author. However, this assumption will be invalid for articles that intentionally list authors in alphabetical order. If some disciplines predominantly use the alphabetical scheme, treating first authors as lead authors will result in a large level of "contamination". In each field, we test for the prevalence of the alphabetical scheme by counting "alphabetical articles" and comparing them to a number expected if the author placement is not intentionally alphabetical. For a paper with $n$ authors we expect $1/n!$ to have alphabetical ordering by chance (for example, 50% in two-authored articles). For astronomy, robotics and ecology this percentage is between 51% and 52%, i.e., these fields generally do not use the alphabetical scheme. On the other hand, in both mathematics and economics this fraction is 97%. Correspondingly, the level of contamination (fraction of authors that would be incorrectly treated as lead authors) is only several percent for astronomy, robotics and ecology, but very high 40% for mathematics and economics. Alternatively, one may treat *corresponding* authors as the lead authors. Using corresponding authors reduces the contamination rate in mathematics and economics to around 24%, which is still relatively high (for example, ~77% of two-authored papers are still alphabetical). Apparently, for mathematics and economics, we need to use a new method that will select only articles in which the lead author can be determined unambiguously. Thus we select: single authored papers, articles in which the corresponding author is not the first author and articles that are not in the alphabetical order.[2] This filtering results in the removal of 41% of articles in mathematics and 46% in economics, which was necessary to obtain a data set with virtually no contamination. We emphasize that this removal does not preferentially affect certain categories of authors and therefore will not lead to biases in the analysis.

---

[2] While it was not essential to do so, we applied the same removal of alphabetical articles (in which the corresponding author is the first author) for astronomy, robotics and ecology, but only for articles with five (four for astronomy) or more authors, because it is among those categories that the alphabetical articles greatly exceed the expected number arising by chance. The result is that the contamination is reduced to <1% with only very small removal of articles.

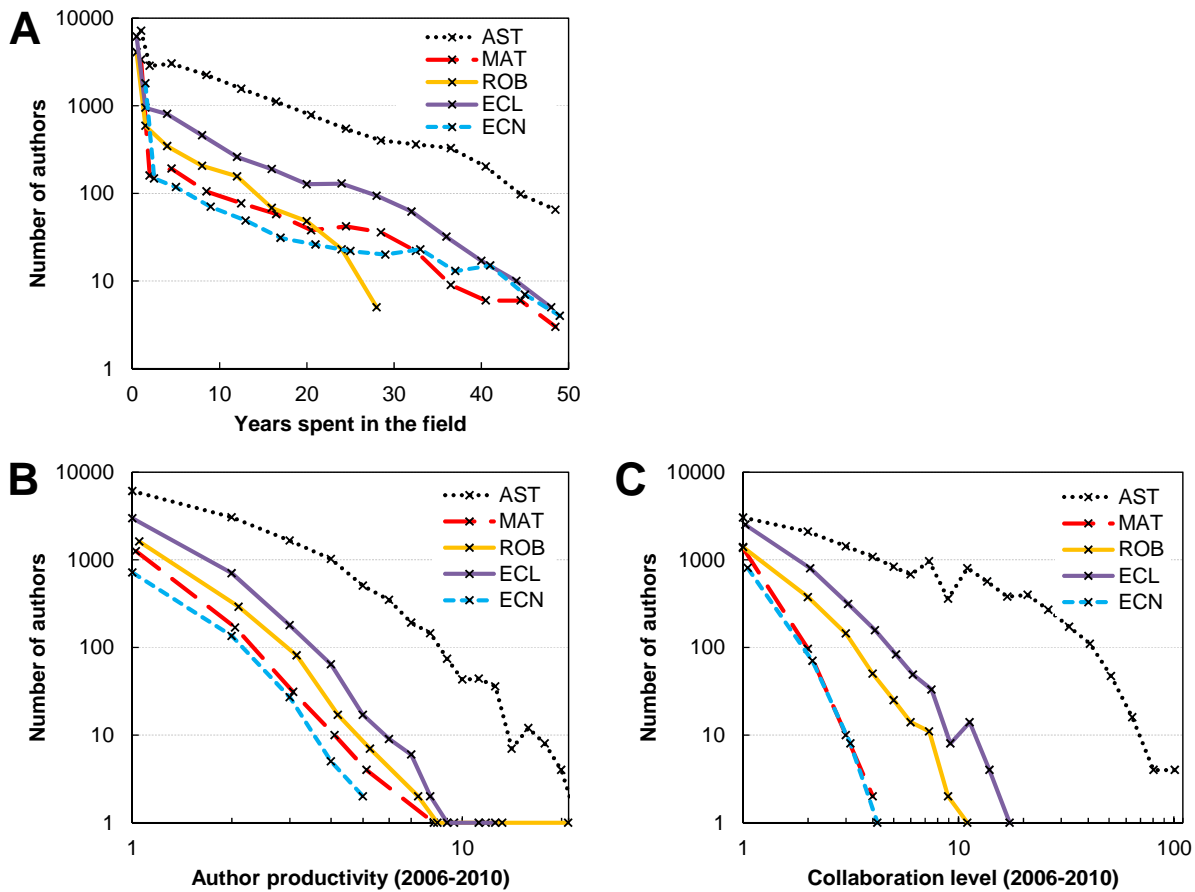

**Figure 1. Distributions of authors of different academic age (A), recent productivity (B), and collaboration level (C). Note the logarithmic scale on the *y* axis and on the *x* axis for productivity and collaboration. Highest bins in every category contain very few authors. Distribution of the academic age (panel A) is per bin, and uses 4-year bins beyond year two. First bin (*x*=1) in panel A is higher for all fields because it contains transient authors who appear with only one publication. Decline is approximately exponential in A and power-law in B and C.**

We identified 21,562 different lead authors in the current period. We disambiguated author names using last names and first (or first and middle) initials. For each lead author in a given field we determined three *independent* measures:

- *Time spent in the field (i.e., academic age)* – span in years, for a given author, between the first and the most recent article in the entire dataset (1961-2010), regardless of authorship placement: $a_{acad} = t_{last} - t_{first} + 1$. Maximum is 50. The academic age distribution drops exponentially, so very few authors are near this maximum (Figure 1A). Operationalization of the academic age through publications is an approximation, but we believe it is a good one, because it represents the length of an author's active engagement within a scientific community
- *Current productivity* – number of lead-author articles published between 2006 and 2010. We excluded co-authored articles not led by an author to remove correlation with the collaboration level. Distribution of author productivity is shown in Figure 1B. It is approximately power-law.

- *Current collaboration level* – number of collaborators from co-authored articles in the current period plus one. Collaborators of a given author are defined as all the different lead authors on articles on which he/she was a coauthor. This measure does not include coauthors on articles on which the author him/herself was the lead author, thus removing correlation with the productivity. Furthermore, we are interested in direct collaborative ties, therefore two authors, neither of whom was the lead author on a multi-authored article, were not considered collaborators. To get the collaboration level we added one in order to enable representation on a logarithmic scale. The distribution of collaboration level is shown in Figure 1C. It roughly follows the power-law distribution (except astronomy). The drop is steepest for mathematics and economics – these are the least collaborative fields. Ecology and robotics are moderately collaborative, while astronomy is highly collaborative (shallowest decline).

For trends involving productivity and collaboration we presented the data using partial logarithmic binning [44] but averaging only non-zero elements. Error bars represent standard deviations of the mean. Logarithmic representation is used in order to show the full dynamic range of author-related quantities more evenly.

## Characteristics of references

For each article we determined the number of references $(N_{all})$ by counting references in the WoS record. We introduced the measure of references per article *page*: $n = N_{all} / p$ to obtain a normalized measure, which we refer to as *reference rate*. Reference rate, by taking into account the expectation that longer papers will have more references, is a more uniform measure of the usage of references than the bulk number of references.

Further, for each article we calculated the following age-related characteristics:

- *Average age of references*:

$$\bar{a} = \frac{1}{N_{all}} \sum (2010 - t_{pub})$$

- *Price Index* – the fraction of references published within five years of the article publication year:

$$PI = N_{\leq 5} / N_{all}$$

- *Modified Price Index* - the fraction of references published within five years compared to those published within 10 years:

$$MPI = N_{\leq 5} / N_{\leq 10}$$

  This measure is similar to the Price Index. However, instead of comparing recent references to all references, it compares recent references only to those published within 10 years. This measure gives a better assessment of the speed of the research front than the traditional Price Index by eliminating the role of "old," foundational references.

- *Old fraction* – fraction of references older than 10 years:

$$O = N_{>10} / N_{all}$$

  This measure can be thought of as the opposite of the Price Index. It indicates the contribution of *foundational* references. Another advantage of the Modified Price Index, compared to the regular Price Index, is that it is completely independent from the old fraction.

Finally, for all current authors who had published two or more papers as lead authors in 2006-2010 we calculated the re-citation fraction at the publication level (the re-use of the same article) as:

$$r = 2\left(1 - \frac{N_{unique(1,2)}}{N_{all,1} + N_{all,2}}\right)$$

Where $N_{unique(1,2)}$ is the number of unique references in two articles and $N_{all,1}$ and $N_{all,2}$ are the total number of references in two articles respectively. If an author had published more than two articles we randomly chose two from which the re-citation fraction was calculated. To increase the accuracy of the re-citation measure we perform the two-article drawing 10 times, thus sampling many different pairs of articles of a given author.

## Results

### Usage and characteristics of references over the last 50 years

The purpose of this section is to provide the context for the rest of the analysis by describing the macro-level characteristics of references: how references differ from field to field, and how their characteristics evolved in each respective field over the last half century.

### Trends in numbers of references

As previous studies have reported, we observe that the number of references per article had been rising in all five fields (Figure 2A). The rate of increase in all fields was similar, with a doubling time of around 30 years. Interestingly, we do not observe any change in the rate of referencing in the late 1990s, when the Internet became more widespread. Ecology and astronomy currently use around 50 references per article on average, twice as many as the other three fields.

The increase in the number of references per article over the period studied can be attributed to two distinct factors: (a) due to articles becoming longer or (b) due to a higher *rate* of referencing. The increase could also result from the combination of these two factors. To distinguish these factors we examined the trends in the number of article pages and in the number of *references per page* (Figure 2B and 2C). Article length has remained remarkably constant in astronomy, robotics and ecology. However, in mathematics and economics there has been a steady increase since at least the 1970s. On the other hand, when it comes to the rate of referencing (number of references per page), astronomy and ecology have seen a steady rise throughout the period under study, robotics has seen an overall rise, while the rate in mathematics and economics remained constant. Therefore, we conclude that different factors contributed to the increase in the number of references per article in these fields. In astronomy, ecology, and robotics this was due to the increase in the rate at which references were being used; in mathematics and economics the articles became significantly longer over time but the number of references per page remained fairly constant. The finding regarding economics agrees with Frandsen and Nicolaisen [26], who found that in the subfield of econometrics "we expect about a 7% increase in the number of references when the number of words increases by 10%" (p. 68). However, it appears that in fields with a strong experimental/observational component there is an increase in reference rate without the increase in the amount of text.

It is interesting that astronomy and ecology articles are the shortest but have the largest number of references, hence their references rate is currently some five times higher than for mathematics, robotics, and economics. As with the distribution of the bulk number of references per paper, we observe no changes in reference rate that would correlate with the spread of the Internet.

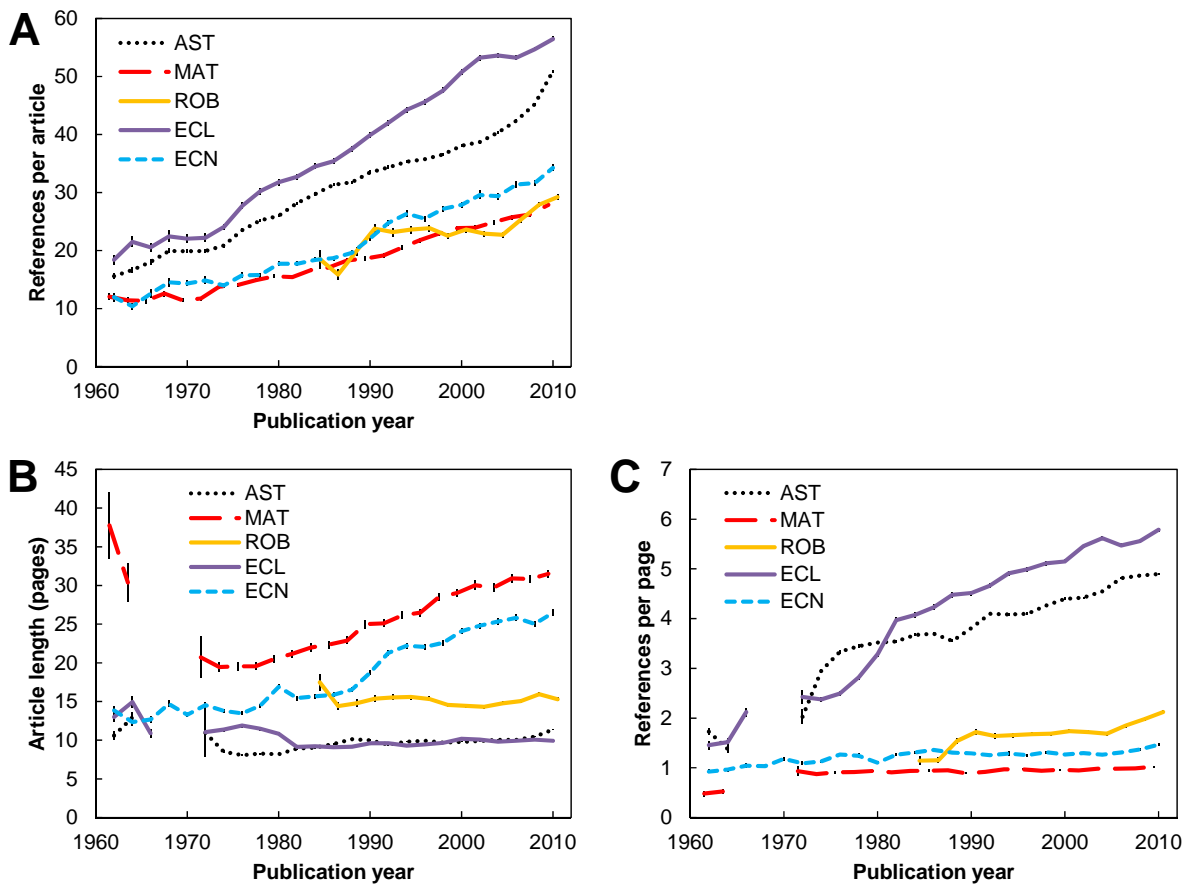

**Figure 2. Trends involving numbers of references for 5 disciplines (astronomy (AST), mathematics (MAT), robotics (ROB), ecology (ECL), and economics (ECN)) from 1961-2010. A) The number of references per article. B) The article length, i.e. number of pages per article. C) Number of references per page (reference rate). Data for ROB start in 1983. Page information is missing for some articles in the late 1960s. Data points are averaged in bins of 2 years.**

### Trends in age-related characteristics of references

Overall, in most fields (to lesser degree in ecology) the average age of references has *risen* since the 1980s (Figure 3A). In astronomy and ecology (and to some degree economics) this period was preceded by one in which the references were on average getting *"younger"* in the 1960s and 1970s. Our analysis partially confirms Larivière and colleagues' [14] finding that "contrary to a widely held belief, the age of cited material has risen continuously since the mid-1960s" . Our data show that this average aging started more recently for fields in this study. Increase in the average age is to some degree a consequence of the aging of the knowledge base, which happens even with the exponential growth in literature that has started at some point in time [35]. However, the *old fraction* (fraction of references older than 10 years, Figure 3B) will be affected only moderately by this "mathematical" aging. If we assume an exponential growth in literature that started in the 1910s then, for typical growth rates, the old fraction should increase by only 0.07 over the period 1960-2010 (based on [35]). We see from Figure 3B that the trends in the old fraction are much larger than 0.07, thereby reflecting real changes in the use of references. Moreover, as with the average age the old fraction was decreasing for astronomy and ecology in the 1960s and 1970s. Overall, the trends in the old fraction generally follow those of average age, meaning that the latter measure is driven primarily by the level of use of foundational references. For mathematics and economics, (the fields that kept the referencing rate but increased article length) this fraction is higher now than it was in the 1960s.

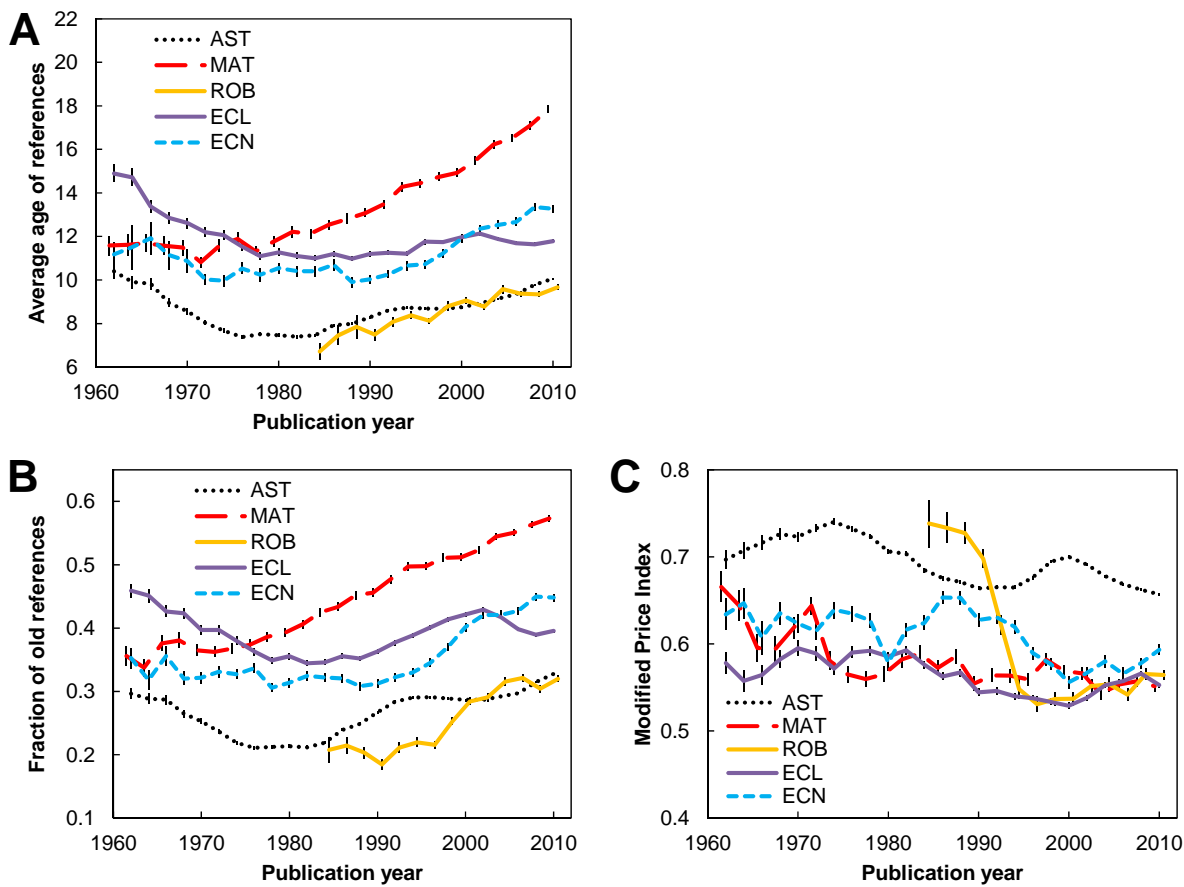

**Figure 3. Trends involving ages of references for five disciplines (astronomy (AST), mathematics (MAT), robotics (ROB), ecology (ECL), and economics (ECN)) from 1961-2010. A) Average age of references. B) Fraction of old references (>10 years old). C) Modified Price Index (fraction of references published within five years compared to those published within 10 years). Data points are averaged in bins of two years.**

How can we explain the trends? If we take the view of Allen and colleagues [45] that the age of the references can be viewed as the age of "persuasive communities" and can therefore be used to reveal the characteristics of scientific communities, then the fields or the time periods characterized by larger fractions of young references represent recent persuasive communities, which are characterized by very rapid change. The change is explained by paradigm shifts that lead to rapid successions of references that are considered acceptable (are able to persuade) by a larger community. On the other hand, the fields that have very old references may be "typified by highly stable, or even tradition-bound science" (p. 301). In that respect, the 1960s and 1970s (we do not have the data for the preceding periods) can be considered revolutionary for astronomy and ecology. Afterwards, these fields can be said to have moved to a more stable/normal phase of development building on the foundational work, although ecology shows some indications of entering another period of lesser use of foundational references. On the other hand, mathematics has been stable through the whole period. With the age of references constantly on the rise, the references in mathematics are now considerably older than they were prior to the 1980s. In other words, mathematics has continued to build on the foundational work that occurred prior to the 1960s. This finding is interesting in the light of what we have found regarding the number of references. While the rate of referencing in mathematics has not changed, its character (age) has. The robotics journals appeared in the early 1980s. However, the age of the references has been more or less constantly on the rise. Although it is the youngest of all fields, robotics has features of fields in their stable phase.

The most commonly used measure of the speed of the research front, the Price Index, is sensitive to old references, the presence of which does not necessarily indicate a slower research front. Thus we introduce a new measure: the Modified Price Index, which by giving the share of recent (<5 years) references compared to those of intermediate age (<10 years) gives a better measure of the actual speed of changes – with higher values meaning higher speed. Figure 3C shows that there is an overall deceleration for all fields, interrupted with periods of acceleration. For robotics we see a huge drop in the 1990s, but some acceleration since then. Astronomy had a temporary boost in the mid-1990s, but is now slowing. Larivière and colleagues [14] also observed this boost in 1990s; they attribute it to the creation of the e-print repository arXiv. However, the prediction that arXiv would lead to the continual lowering in the age of citations did not come to fruition; the change was only temporary. Economics had a period of accelerated research front in the 1980s. As indicated previously, in the current period it is only ecology that exhibits some signs of accelerated development.

In the current period, astronomy and robotics references remain the youngest (~10 years), mathematics references are the oldest (~18 years), while both ecology and economics are in between (13-14 years). These differences in age reflect mainly the differences in the fraction of foundational references (30% for astronomy and ecology to 60% in mathematics).On the other hand, when it comes to the Modified Price Index all fields are very similar except astronomy, whose research front moves faster than that of other fields.

The conclusion from this section is that fields, even with fast moving research fronts, will show aging in their references whenever they are in a stable phase, building on foundational work. It is only during the periods of paradigm shifts (even for classical fields) that this trend will be reversed. Currently, all the disciplines, even robotics, are in the stable, "traditional" stages of development.

## Referencing behavior of authors

We now turn to the meso-level analysis of referencing behavior: the relationship between particular characteristics of an author (academic age, productivity, and collaboration) and his/her referencing behavior (rate of reference usage, age characteristics of references, and re-citation rates). To remove the dependence from global trends discussed above, from this point on we consider only the literature from the most recent five-year period (2006-2010).

### Dependence on academic age

Previous studies of referencing behavior focused on the actual age of researchers and have suggested the following trends: (a) the more senior the author, the fewer references he/she uses, (b) those references are older, and (c) he/she tends to re-cite more. We focus on the academic age as a probably more relevant type of age (the two will, of course, be to large degree related).

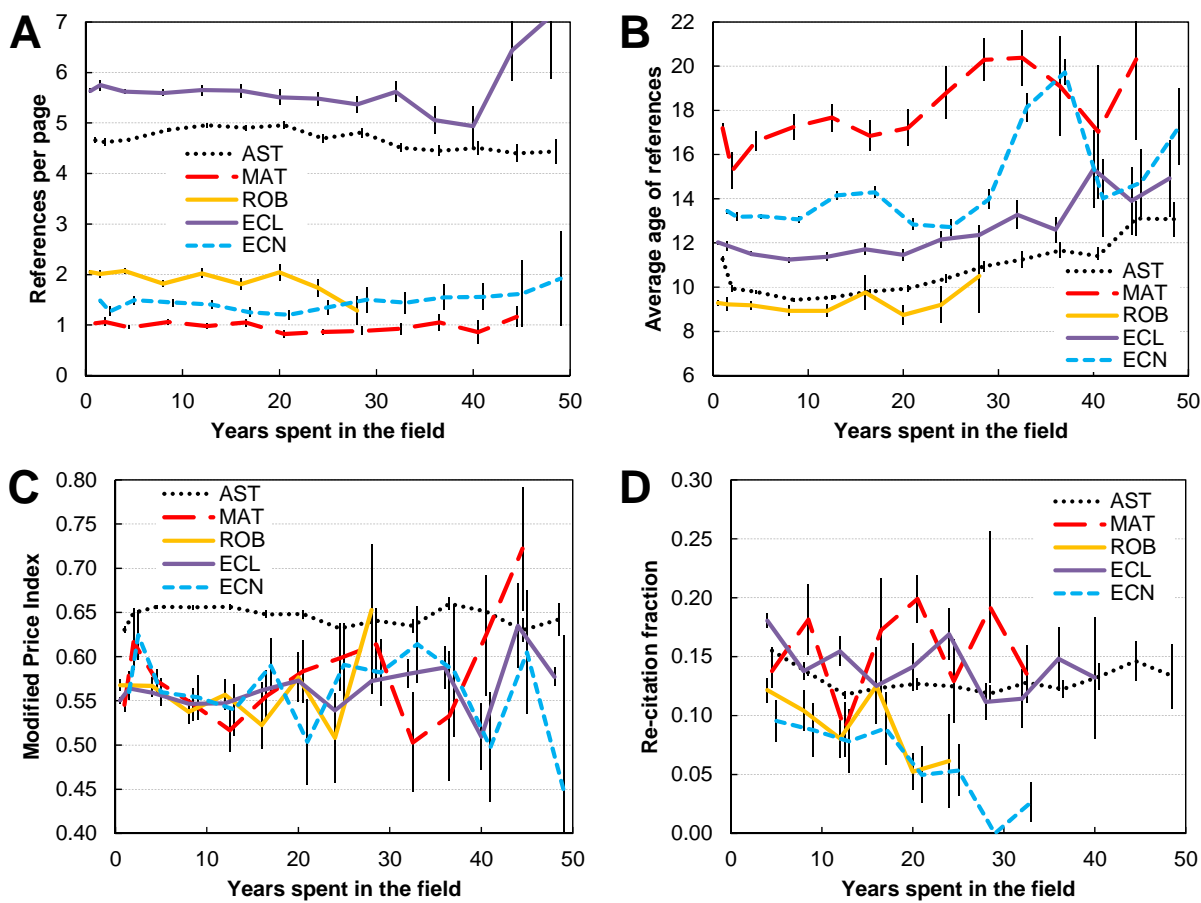

Figure 4. Referencing behavior for authors of different academic age (number of years spent in the field) in five disciplines (astronomy (AST), mathematics (MAT), robotics (ROB), ecology (ECL), and economics (ECN)). A) Number of references per page (reference rate). B) Average age of references. C) Modified Price Index (fraction of references published within five years compared to those published within 10 years). D) Re-citation fraction (repeated citations in pairs of articles). Authors who appear once have $x = 1$. First two years not binned; afterwards data are averaged in bins of four. Bins with fewer than 4 authors (rightmost, see Figure 1A) are excluded. For re-citation (D) points are not shown for the first two years because the average span between the pair of publications is less than it is for later ages.

To control for differences in article lengths, we analyzed the relationship between the academic age and number of references per page (reference rate). We find (Figure 4A) that the notion that senior authors use fewer references is generally not true in any of the five fields studied here. In astronomy, ecology and robotics there is a very slight downward trend (total decrease of ~10%) especially after 15 to 20 years spent in the field. In mathematics and economics there is no statistically significant trend[3] at all. Although the reference rate in astronomy and ecology today is much higher than what it was when academically older authors were establishing their practices (Figure 2C), these authors have kept up with the times and use the references at the same rate as their junior colleagues.

On the other hand, we confirm the findings of previous studies that senior authors on average use older references (Figure 4B). However, this trend in the mean age is almost entirely due to the increase in the fraction of old, "foundational" references (older than 10 years) (graph not shown), and not due to less innovative research, as we will see shortly. The senior researchers, having spent more time in the field, have more knowledge of the older literature (having had a chance to accumulate their knowledge base longer) and tend to use it more. The exception is robotics, where the trend is not statistically significant with the current data, but note that for robotics we do not have the data beyond three decades. In most fields the difference in reference ages starts becoming pronounced for scientists active for more than a decade. The "youngest" references are used by the scientists who have been active for around a decade, which does not necessarily need to be interpreted as a sign that these researchers are at the forefront of research [13,32] (as we will confirm shortly). While they may be the ones who are the least burdened by either tradition or authority, they are also likely less familiar with the foundational work, and therefore cite it less. The age of references is only one side of the coin. In order to test whether authors who have spent long time periods in the field actually lose an "edge," as suggested in previous studies, we calculated their Modified Price Index. Gingras and colleagues [13] found that the young authors (between 28 and 40) have the highest *regular* Price Index, with the index steadily falling afterward. However, the regular Price Index, unlike our modified index, does not isolate foundational references. As can be seen in Figure 4C, all fields except astronomy show strong fluctuations in the Modified Price Index, but are generally consistent with no change. Thus, although senior researchers tend to have a higher percentage of old references, they don't fall behind their junior colleagues when it comes to citing the most recent literature.

Finally, it has also been suggested [13] that researchers actively follow the literature and accumulate references until they are about 40 and reuse their accumulated sets of references after that. This would translate into higher re-citation fractions for senior researchers. We show trends of re-citation rate for authors of different academic age in Figure 4D. In terms of the overall re-citation level, astronomy, mathematics and ecology all have similar values (13%). Robotics is somewhat lower (11%), and economics is significantly lower (7%). Again, our results do not fit the expected picture. The re-citation rates are either consistent with no overall change (astronomy and mathematics), or are even dropping (for the other three fields). In astronomy, where the data are of higher quality (many more authors with more than a single publication), we see that the re-citation rates are slightly higher at both the beginning of their career (<10 years) and towards the end (>35 years). However, overall there is no trend that would associate older scientists with much higher re-citation rate.

While the focus of this section was mostly on older vs. younger authors, let us now focus on the authors who have just entered the field (shown at $x = 1$, the first bin in Figure 4ABC), a category that also contains a large fraction of authors who will not continue to publish (the transients, see Figure 1A). Interestingly, in all the fields except robotics the newcomers/transients tend to use somewhat older references and have a lower Modified Price Index. Such referencing behavior is to some extent because the entering researchers still depend heavily on their acceptance from the gatekeepers (who by default tend to be senior and thus have a proclivity for somewhat older references), but more likely because they need to build trust and authority by showing their knowledge of the foundations of the field. For example, many of those works are probably derived from dissertations, where the review of foundational work is more important than in regular science papers. Also, the papers based on dissertations are apparently often not at the research forefront, reflecting research conducted over a number of years and so not always including the most up-to-date references.

---

[3] In this study we discuss only the trends that are statistically significant for a given range in $x$, that is, when the probability of the linear trend (with error bars in $y$ taken into account) arising by chance is smaller than 5%.

To summarize, analyzing the referencing behavior of authors with different academic ages led to the shattering of some of the myths regarding the practices of senior authors. Overall, we find that they use references at the same rate as their junior colleagues, with the same rate of re-citation. Although they tend to use a higher fraction of foundational references, their Modified Price Index indicates that their research does not lack very recent references. On the other hand, for authors who have just entered the field, many of whom will not continue in research careers, their MPI reflects less innovative work.

### Dependence on current productivity

Do prolific authors tend to economize on references, or does their increased productivity also expose them to a larger body of literature that they cite? Previous studies have not focused on this factor. We show results in Figure 5A. For mathematics, robotics and economics neither is the case – productivity has no effect on the referencing rate. There is also no difference in the article length, so the total number of references is also constant (graphs not shown). In ecology and astronomy, on the other hand, more prolific authors have significantly *higher* referencing rates.

Figure 5B shows that more prolific authors tend to use on average "younger" references (trends are significant in astronomy, robotics and ecology). This finding may be used to support the idea that, at least in some fields, it is the more productive authors whose research is farthest from the foundational work in their field. To further test this idea, we show the trends regarding the relationship between author productivity and Modified Price Index (Figure 5C). We observe that researchers in all fields except robotics show a significant increase in the Modified Price Index as their productivity increases. This can further support our findings that the productivity (as measured here by the number of lead author publications) is on average positively related to the quality of one's research.

Figure 5D shows the relationship between re-citation and the author's productivity. If the very productive authors were to continue to work on the same topic we would expect high re-citation fractions. However, that is not the case in any of these fields. On the contrary, there is a hint of an opposite trend, which is statistically significant for astronomy (where the data are most reliable) in that we observe that the re-citation fraction actually decreases for more productive authors. Recall that the re-citation fraction is determined only from articles led by the author, so the lower re-citation rate is not an artifact of co-authorship.

Based on these analyses it is fairly safe to say that for most of the fields the most productive authors are the ones engaged in the most innovative and varied research, as evidenced by their usage of more recent literature and an increase in the Modified Price Index. In addition, these researchers draw from a larger pool of knowledge, indicating that they may be moving more often from topic to topic. These findings can be used to dispel the notions of most prolific authors engaging in the so-called "salami" publishing, at least for authors who publish their numerous papers in high-quality *core* journals.

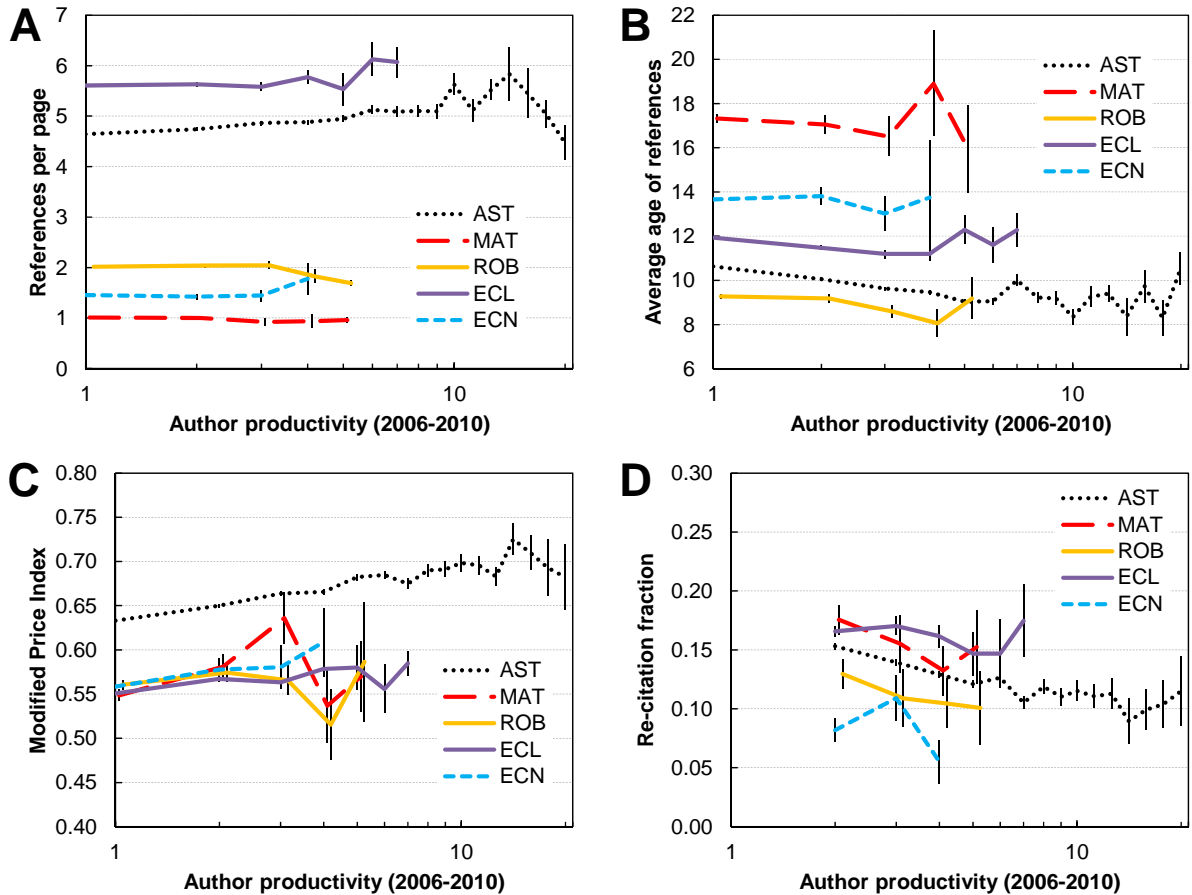

Figure 5. Referencing behavior for authors of different recent productivity in five disciplines (astronomy (AST), mathematics (MAT), robotics (ROB), ecology (ECL), and economics (ECN)). A) Number of references per page (reference rate). B) Average age of references. C) Modified Price Index (fraction of references published within five years compared to those published within 10 years). D) Re-citation fraction (repeated citations in pairs of articles). Data are binned in intervals of 0.05 decades. Bins with fewer than four elements are excluded.

### Dependence on current level of collaboration

Do highly collaborative authors tend to use references differently from authors in their field who work alone or with very few people? We define collaboration through a number of recent papers to which an author contributed as a co-author, thus decoupling it entirely from the productivity (papers which the author led). Before describing the trends we want to say a few words regarding collaborative practices. In all fields studied here except astronomy most researchers have few or no collaborators (Figure 1C), which we associate with a "normal mode" of collaboration [46], a mode that is not a result of the rich-get-richer phenomenon (preferential attachment mode) or from large-team projects (hyperauthorship mode). Only in astronomy can we follow the referencing trend for authors who have many collaborators (up to ~70) and are involved in the "big science" aspect of this field.

Generally, we find that authors of various collaborative affinities within the normal mode use references in their works at the same rate. For authors in astronomy who collaborated with more than four researchers the referencing rate increases steadily continuing well into preferential attachment mode. It is likely that the collaborative activity above some threshold exposes an author to a wider body of literature, which the author then uses in his/her publications.

We also found a universal trend (significant in all fields except economics) that the authors who collaborated more used, on average, "younger" references (Figure 6B). The trend was especially pronounced for astronomy, where there was a very marked difference between solo authors and those with ~50 collaborators (11.6 vs. 6.9 years on average). Thus, we may tentatively conclude that the authors who were more collaborative dwell the least on foundational work and push the frontiers of research. To test whether the latter is true we examine the Modified Price Index.

Similar "benefits" of collaboration are evident in the trends of the Modified Price Index (Figure 6C). In astronomy, authors who collaborated more used a higher fraction of recent references within the body of references up to 10 years old. The fact that the MPI was especially high for authors in astronomy who participated in "big science" projects supports the argument that scientists who participate in big projects, *and publish as lead authors,* are those who push the frontiers of the field. Trends are present, but weaker in less collaborative fields such as ecology and robotics. In fields where collaboration is rare, like mathematics and economics, those who do collaborate have a similar MPI to those who do not.

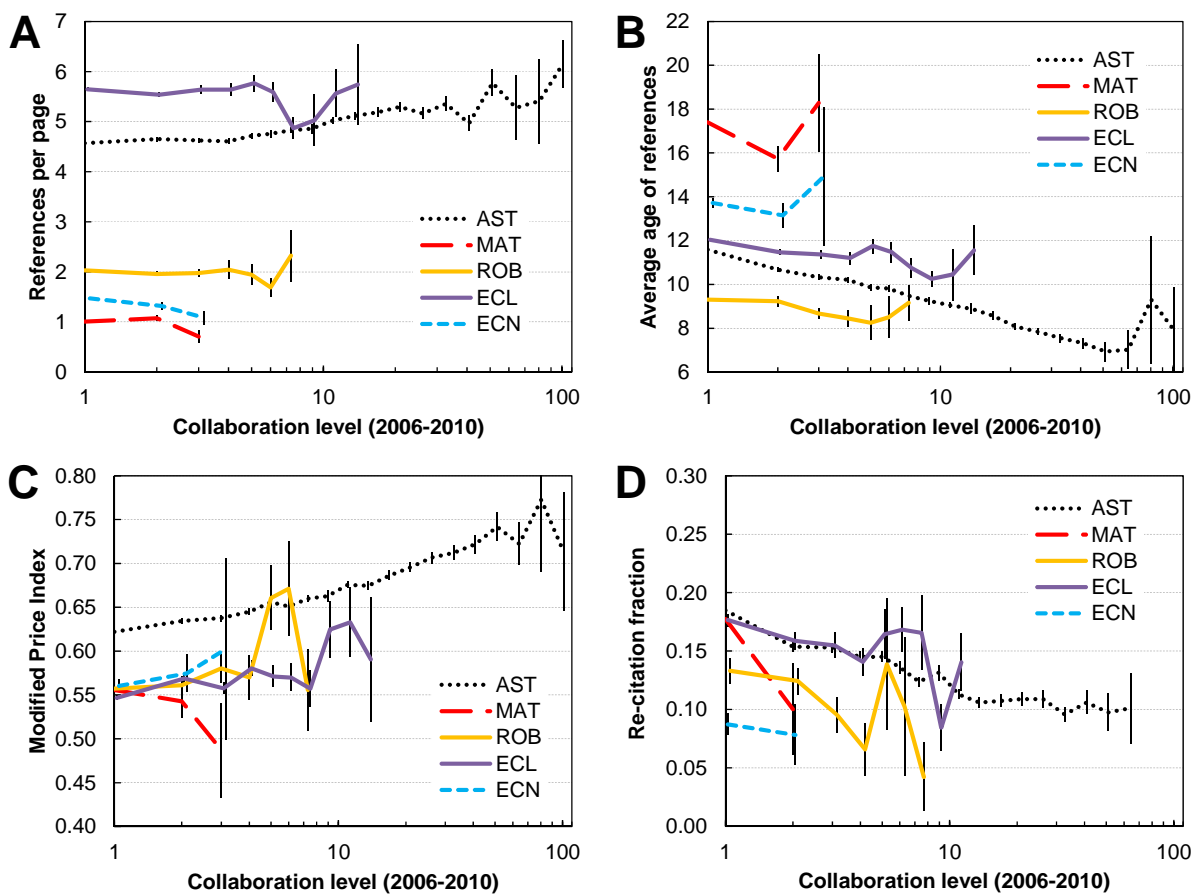

**Figure 6. Referencing behavior for authors of different recent collaboration level in five disciplines (astronomy (AST), mathematics (MAT), robotics (ROB), ecology (ECL), and economics (ECN)). Authors with no collaborators are shown at collaboration level 1, those with one at *x*=2, etc. A) Number of references per page (reference rate). B) Average age of references. C) Modified Price Index (fraction of references published within five years compared to those published within 10 years). D) Re-citation fraction (repeated citations in pairs of articles). Data are binned in intervals of 0.1 decades. Bins with fewer than four elements are excluded.**

Lastly, we find that authors who collaborate more tend to re-cite less. In all fields, except economics, we observe a very strong decline of re-citation (Figure 6D). Thus, collaborative activity appears to go hand in hand with more varied research.

The trends in this section are qualitatively very similar to trends with respect to productivity. Like productive authors, those who were very collaborative tended to do most innovative science based on the younger references. In addition, these authors were more versatile in terms of the topics they covered, as evidenced by the decrease in their re-citation fraction. We emphasize that our measures of productivity and collaboration are completely independent in that they are based on different sets of documents (papers led by an author vs. papers not led by an author). Yet, the two may be correlated due to a common cause. Indeed, we verify that the mean productivity and collaboration are correlated in all fields. It may not be surprising that the researchers who have the most to offer to others (and are therefore being invited to join collaborations led by others) are the ones who themselves produce the most.

## Discussion

### Relative impact of author-related factors on the citing behavior

The results presented in the previous section explored how a certain characteristic of a group of authors in a given discipline – the academic age, productivity or collaboration level – correlates with the references that those authors used.

For most fields, the rate at which authors currently use references (references per page) is very weakly dependent on any of the three author characteristics. The exception is astronomy, where all three factors lead to significant trends – the most important being the increase of the referencing rate for authors who collaborated more.

Author-related characteristics are much more relevant for the age of references used. The strongest trend was that authors who collaborated more used references with significantly lower average age (typically by two years, but in astronomy the effect was up to five years). References that were, on average, "younger" by several years were also characteristic of authors who recently entered the field (but not those who had just entered). Finally, more prolific authors also tend to use references that are up to two years "younger". There are also some field-specific differences. For example, in astronomy and ecology, collaboration has a somewhat larger effect than age, while the opposite is true in mathematics and economy, fields in which the authors do not collaborate extensively.

Trends involving the Modified Price Index were similar but not identical to those for the age of references. Active collaboration activity led to strong increases in the index (i.e., their research front moved faster) in astronomy, robotics and ecology. No author characteristic is strongly correlated with the MPI in economics. There was also some increase of the index for authors who were more productive. Interestingly, academic age was not a strong factor in determining the Modified Price Index – senior authors typically had similar indexes to those who had recently entered the field.

Re-citation fractions, which we interpret as being negatively correlated with the variety of topics on which scientists work, either were not dependent on the time spent in the field or actually went down for more senior researchers. For astronomy, there was some negative correlation between the re-citation fraction and productivity. However, the strongest correlation is with the number of collaborators.

Speaking generally, we conclude that the collaboration level is probably the most important factor related to citing behavior, especially in fields with extensive collaborative practices.

## Possible age-related biases

Our results indicate that, in general, more senior authors do not fall behind their junior colleagues. So, is the picture of an out-of-touch older scientist entirely wrong? Obviously one can only study citing behavior of authors who publish. Furthermore, our study analyzes references in core journals, which have higher impacts. If older scientists preferentially gravitate towards lesser impact journals and if their citing behavior is rather different from those who still publish in core journals then our study will not necessarily represent their citing behavior. Whether that is actually the case is outside of the scope of this study. Rather than being a limitation of our study, the analysis only of authors who publish in core journals simply means that we explore the behavior of that section of the scientific community which is the most active in pushing the scientific frontiers.

# Conclusions

It has been claimed that paradigms, research traditions and disciplines provide a framework or guidelines for problems and their possible solutions. Do they also provide frameworks on both how and what we reference, which is an essential component of reporting results in research articles and therefore of scientific communication? Do other factors, such as the traits of individual authors also play a role?

The properties of references in a given article, such as their number, average age, or fraction of recent references are the result of referencing behavior and can tell us about the character of research. In this study we investigated referencing behavior in five diverse fields (astronomy, mathematics, robotics, ecology and economics) based on 213,756 articles published in core journals of their respective fields. The study first revisited the referencing behavior at the macro level, following the trends of reference usage in the five fields over the last 50 years. We confirmed earlier results that both the discipline and the time period are strong determinants of citing behavior. We found steady increases in the number of references, but found that it arises from diverse causes: in some fields it was due to a higher rate of usage (number of references per page), while in others it reflected longer articles at more recent times. In all fields the fraction of references older than 10 years had been increasing since the 1980s beyond what is expected from the pure mathematical aging in the exponentially growing body of literature. Contrary to some expectations, the introduction of the Internet apparently resulted in no long-term changes in the referencing behavior.

Our main focus and the novel aspect of this study was to explore current referencing behavior in five fields at the meso level in order to investigate whether different categories of authors use references differently. We analyzed 21,562 authors of differing academic age, productivity and collaboration levels, who published in relatively high-impact core journals, i.e., those who define the currents of modern science. Our results may not necessarily reflect the practices of the entire population of researchers in a given field. Contrary to some previous findings and expectations, we found that senior researchers (who still published in core journals) used references at the same rate as their junior colleagues, with similar rates of re-citation. References of senior authors contained a higher fraction of older foundational works, however their fraction of new (< 5 years) vs. recent (<10 years) references (the Modified Price Index) indicated that their research had a similar cutting-edge aspect. Interestingly, this was less evident for researchers who had just entered the field. We found that both the productive researchers and especially those who collaborated more extensively used a significantly lower fraction of foundational references, had a much higher Modified Price Index and used the same references less repeatedly. These author-related trends were similar in all five fields. In other words, highly collaborative and highly productive scientists (regardless of age) typically possess those characteristics as part of their overall excellence, which is also reflected in the way they use references, which suggests that they are the scientists pushing the research front.

## Acknowledgement

I thank the referee for useful suggestions, Deborah Shaw for careful reading and comments on the manuscript and Colleen Martin for excellent copy editing.

## References

1. Bazerman C (1988) Shaping written knowledge: The genre and activity of the experimental article in science. Madison: The University of Wisconsin Press.
2. Hyland K (2004) Disciplinary discourses: Social interactions in academic writing. Ann Arbor: The University of Michigan Press.
3. Shapin S (1994) A social history of truth: Civility and science in seventeenth-century England. Chicago: The University of Chicago Press.
4. Börner K (2010) Atlas of science: Visualizing what we know. Cambridge: MIT Press.
5. Leydesdorff L, Wouters P (1999) Between texts and contexts: Advances in theories of citation? Scientometrics 44: 169-182.
6. Gilbert GN (1977) Referencing as persuasion. Social Studies of Science 7: 113-122.
7. Kaplan N (1965) The norms of citation behavior: Prolegomena to the footnote. American Documentation 16: 179-184.
8. Small H (1978) Cited documents as concept symbols. Social Studies of Science 8: 327-340.
9. Brooks TA (1985) Private acts and public objects: An investigation of citer motivations. Journal of the American Society for Information Science 36: 223-229.
10. Chubin DE, Moitra SD (1975) Content analysis of references: Adjunct or alternative to citation counting? Social Studies of Science 5: 423-441.
11. Cronin B (1981) The need for a theory of citing. Journal of Documentation 37: 16-24.
12. Barnett GA, Fink EL (2008) Impact of the internet and scholar age distribution on academic citation age. Journal of the American Society for Information Science and Technology 59: 526-534.
13. Gingras Y, Larivière V, Macaluso B, Robitaille J-P (2008) The effects of aging on researchers' publication and citation patterns. PLoS ONE 3: e4048.
14. Larivière V, Archambault É, Gingras Y (2008) Long-term variations in the aging of scientific literature: From exponential growth to steady-state science (1900-2004). Journal of the American Society for Information Science and Technology 59: 288-296.
15. Glänzel W, Schoepflin U (1995) A bibliometric study on ageing and reception processes of scientific literature. Journal of Information Science 21: 37-54.

16. Line MB (1970) The half-life of periodical literature: Apparent and real obsolescence. Journal of Documentation 26: 46-54.
17. Glänzel W, Schoepflin U (1999) A bibliometric study of reference literature in the sciences and social sciences. Information Processing and Management 35: 31-44.
18. Moed HF, van Leeuwen TN, Reedijk J (1998) A new classification system to describe the ageing of scientific journals and their impact factors. Journal of Documentation 54: 387-419.
19. Ajiferuke I, Lu K, Wolfram D (2011) Who are the research disciples of an author? Examining publication recitation and oeuvre citation exhaustivity. Journal of Informetrics 5: 292-302.
20. Wouters P (1999) The citation culture. Unpublished doctoral dissertation, University of Amsterdam.
21. Cole JR (2000) A short history of the use of citations as a measure of the impact of scientific and scholarly work. In: Cronin B, Atkins HB, editors. The web of knowledge: A festschrift in honor of Eugene Garfield. Metford, NJ: Information Today. pp. 281-300.
22. Cronin B (1984) The citation process: The role and significance of citations in scientific communication. London: Taylor Graham.
23. Moravcsik MJ, Murugesan P (1975) Some results on the function and quality of citations. Social Studies of Science 5: 86-92.
24. Greenberg SA (2009) How citation distortions create unfounded authority: Analysis of a citation network. British Medical Journal 339: b2680.
25. Zuckerman H, Merton RK (1973) Age, aging, and age structure in science. In: Merton RK, editor. The sociology of science: Theoretical and empirical investigations. Chicago: The University of Chicago Press. pp. 497-559.
26. Frandsen TF, Nicolaisen J (2012) Effects of academic experience and prestige on researchers' citing behavior. Journal of the American Society for Information Science and Technology 63: 64-71.
27. Joy S (2006) What should I be doing, and where are they doing it? Scholarly productivity of academic psychologists. Perspectives on Psychological Science 1: 346-364.
28. Wuchty S, Jones BF, Uzzi B (2007) The increasing dominance of teams in production of knowledge. Science 316: 1036-1039.
29. Jones BF, Wuchty S, Uzzi B (2008) Multi-university research teams: Shifting impact, geography, and stratification in science. Science 322.
30. Price DJdS (1963) Little science, big science. New York: Columbia University Press.
31. van Raan AFJ (2000) On growth, ageing, and fractal differentiation of science. Scientometrics 47: 347-362.
32. Price DJdS (1970) Citation measures of hard science, soft science, technology, and nonscience. In: Nelson CE, Pollock DK, editors. Communication among scientists and engineers. Lexington: Heath Lexington Books. pp. 3-22.
33. Price DJdS (1965) Networks of scientific papers. Science 149: 510-515.
34. Burton RE, Kebler RW (1960) The half-life of some scientific and technical literatures. American Documentation 11: 18-22.
35. Egghe L (2010) A model showing the increase in time of the average and median reference age and the decrease in time of the Price Index. Scientometrics 82: 243-248.
36. White HD (2001) Authors as citers over time. Journal of the American Society for Information Science and Technology 52: 87-108.
37. Cronin B, Shaw D (2002) Identity-creators and image-makers: Using citation analysis and thick description to put authors in their place. Scientometrics 54: 31-49.
38. Small H (2010) Referencing through history: How the analysis of landmark scholarly texts can inform citation theory. Research Evaluation 19: 185-193.

39. Minguillo D (2010) Toward a new way of mapping scientific fields: Authors' competence for publishing in scholarly journals. Journal of the American Society for Information Science and Technology 61: 772-786.
40. Henneken EA, Kurtz MJ, Guenther E, Accomazzi A, Grant CS, et al. (2007) E-prints and journal articles in astronomy: A productive co-existence. Learned Publishing 20: 16-22.
41. Garfield E (1982) Journal citation studies, 36. Pure and applied mathematics journals: What they cite and vice versa. Current Contents 15: 5-13.
42. Nobis M, Wohlgemuth T (2004) Trend words in ecological core journals over the last 25 years (1978-2002). Oikos 106: 411-421.
43. Liner GH (2002) Core journals in economics. Economic Inquiry 40: 138-145.
44. Milojević S (2010) Power-law distributions in information science - Making the case for logarithmic binning. Journal of the American Society for Information Science and Technology 61: 2417-2425.
45. Allen B, Qin J, Lancaster FW (1994) Persuasive communities: A longitudinal analysis of references in the Philosophical Transactions of the Royal Society, 1665-1990. Social Studies of Science 24: 279-310.
46. Milojević S (2010) Modes of collaboration in modern science - Beyond power laws and preferential attachment. Journal of the American Society for Information Science and Technology 61: 1410-1423.

**Table 1. Core journals in five fields used in this study.**

| Field | Journal title | Articles (1961-2010) | Articles (with references, 2006-2010) | Average number of ref (2006-2010) |
|---|---|---|---|---|
| AST | ASTRONOMICAL JOURNAL | 14392 | 2022 | 48.6 |
| AST | ASTRONOMY & ASTROPHYSICS | 44050 | 9109 | 43.8 |
| AST | ASTROPHYSICAL JOURNAL | 70661 | 12297 | 49.3 |
| AST | MONTHLY NOTICES OF THE ROYAL ASTRONOMICAL SOCIETY | 27768 | 8042 | 47.2 |
| MAT | ACTA MATHEMATICA | 667 | 56 | 35.5 |
| MAT | AMERICAN JOURNAL OF MATHEMATICS | 2515 | 253 | 27.2 |
| MAT | ANNALS OF MATHEMATICS | 2289 | 310 | 30.5 |
| MAT | INVENTIONES MATHEMATICAE | 3477 | 325 | 31.7 |
| MAT | JOURNAL OF FUNCTIONAL ANALYSIS | 5175 | 1301 | 25.0 |
| MAT | JOURNAL OF THE AMERICAN MATHEMATICAL SOCIETY | 513 | 171 | 30.9 |
| MAT | MATHEMATICS OF COMPUTATION | 4560 | 563 | 23.2 |
| MAT | PROCEEDINGS OF THE LONDON MATHEMATICAL SOCIETY | 2406 | 264 | 28.2 |
| ROB | ADVANCED ROBOTICS | 1009 | 424 | 22.3 |
| ROB | AUTONOMOUS ROBOTS | 454 | 208 | 33.8 |
| ROB | IEEE ROBOTICS & AUTOMATION MAGAZINE | 406 | 167 | 23.2 |
| ROB | IEEE TRANSACTIONS ON ROBOTICS* | 1720 | 447 | 30.9 |
| ROB | INDUSTRIAL ROBOT-AN INTERNATIONAL JOURNAL | 409 | 198 | 18.2 |
| ROB | INTERNATIONAL JOURNAL OF ROBOTICS RESEARCH | 1214 | 251 | 35.9 |
| ROB | JOURNAL OF FIELD ROBOTICS** | 1178 | 191 | 31.6 |
| ROB | JOURNAL OF INTELLIGENT & ROBOTIC SYSTEMS | 1093 | 333 | 26.4 |
| ROB | ROBOTICS AND AUTONOMOUS SYSTEMS | 867 | 344 | 29.7 |
| ECL | ECOLOGY | 9753 | 1662 | 50.4 |
| ECL | JOURNAL OF ANIMAL ECOLOGY | 3532 | 661 | 54.0 |
| ECL | JOURNAL OF ECOLOGY | 3627 | 596 | 56.5 |
| ECL | OECOLOGIA | 9990 | 1454 | 54.3 |
| ECL | OIKOS | 6150 | 1043 | 50.8 |
| ECN | AMERICAN ECONOMIC REVIEW | 4502 | 475 | 36.1 |
| ECN | ECONOMETRICA | 2907 | 270 | 32.0 |
| ECN | JOURNAL OF ECONOMIC THEORY | 2857 | 578 | 24.4 |
| ECN | JOURNAL OF MONETARY ECONOMICS | 1731 | 427 | 30.1 |
| ECN | JOURNAL OF POLITICAL ECONOMY | 2700 | 158 | 36.9 |
| ECN | QUARTERLY JOURNAL OF ECONOMICS | 2080 | 206 | 42.4 |
| ECN | REVIEW OF ECONOMIC STUDIES | 2030 | 237 | 37.6 |
| | Total | 213756 | 45043 | |

\* Includes: IEEE JOURNAL OF ROBOTICS AND AUTOMATION and IEEE TRANSACTIONS ON ROBOTICS AND AUTOMATION

\*\* Includes: JOURNAL OF ROBOTIC SYSTEMS

Mean number of references per article in each journal is given in the last column. Field abbreviations: astronomy (AST), mathematics (MAT), robotics (ROB), ecology (ECL), and economics (ECN).